\documentclass[reprint,superscriptaddress,amssymb, amsmath, aps, showpacs, footinbib,  prl]{revtex4-1}
\usepackage{graphicx}

\newcommand{\gap}{\text{gap}}
\newcommand{\LMTO}{\text{LMTO}}
\newcommand{\eps}{\varepsilon}

\begin{document}

\title{Orbital order and magnetism of FeNCN}
\author{Alexander A. Tsirlin}
\email{altsirlin@gmail.com}
\affiliation{Max Planck Institute for Chemical Physics of Solids, N\"{o}thnitzer
Str. 40, 01187 Dresden, Germany}

\author{Klaus Koepernik}
\affiliation{IFW Dresden e.V., P.O. Box 270116, D-01171 Dresden, Germany}

\author{Helge Rosner}
\email{Helge.Rosner@cpfs.mpg.de}
\affiliation{Max Planck Institute for Chemical Physics of Solids, N\"{o}thnitzer
Str. 40, 01187 Dresden, Germany}

\begin{abstract}
Based on density functional calculations, we report on the orbital order and microscopic magnetic model of FeNCN, a prototype compound for orbital-only models. Despite having a similar energy scale, the spin and orbital degrees of freedom in FeNCN are only weakly coupled. The ground-state configuration features the doubly occupied $d_{3z^2-r^2}$ ($a_{1g}$) orbital and four singly-occupied $d$ orbitals resulting in the spin $S=2$ on the Fe$^{+2}$ atoms, whereas alternative ($E_g'$) configurations are about 75~meV/f.u. higher in energy. Calculated exchange couplings and band gap are in good agreement with the available experimental data. Experimental effects arising from possible orbital excitations are discussed.
\end{abstract}

\maketitle

The relationship between spin and orbital degrees of freedom in transition-metal compounds is well established on both phenomenological and microscopic levels. Phenomenologically, the orbital states are described by pseudospin operators and can be treated using diverse techniques developed for spin Hamiltonians~\cite{kugel1973,*kugel1982}. Microscopically, the orbital pattern determines the superexchange couplings that are the main driving force of magnetism in insulators. The opposite effect, the influence of magnetism on the orbital state, is less universal~\cite{mostovoy2004} than initially expected~\cite{kugel1973,*kugel1982}. Particularly, recent computational studies of model orbitally-ordered materials~\cite{pavarini2008,*pavarini2010} indicated the important role of lattice distortions in stabilizing specific orbital states. Although spin and orbital degrees of freedom are inherently tangled, there has been considerable theoretical effort in exploring orbital-only models with no spin variables involved~\cite{brink2004,rynbach2010,wenzel2011}. In this paper, we will present a compound that features intrinsically weak coupling between spins and orbitals, and may be a feasible experimental probe for such orbital-only models.

\begin{figure}
\includegraphics{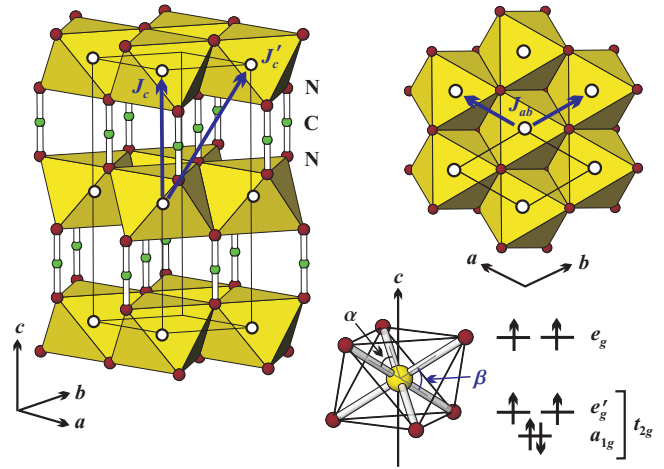}
\caption{\label{fig:str}
(Color online) Left panel: crystal structure of FeNCN. Right panel: close-packed layer of FeN$_6$ octahedra (top), the single octahedron squeezed along the three-fold $c$ axis (bottom left), and a sketch of the ground-state $d^6$ electronic configuration of Fe$^{+2}$ (bottom right). 
}
\end{figure}

As a model compound, we consider the recently discovered iron carbodiimide FeNCN containing Fe$^{+2}$ cations that form layers of close-packed FeN$_6$ octahedra in the $ab$ plane~\cite{liu2009}. The layers are connected via linear NCN units (Fig.~\ref{fig:str}). Experimental information on FeNCN is rather scarce. The compound is a colored (dark-red~\cite{liu2009} or brown~\cite{xiang2010}) antiferromagnetic insulator with the N\'eel temperature of 345~K~\cite{liu2009}. The electronic structure of FeNCN was studied by Xiang \textit{et al.}~\cite{xiang2010} who arrived at a puzzling conclusion on the dramatic failure of conventional density functional theory (DFT)+$U$ methods that were unable to reproduce the experimentally observed insulating ground state. We will show that this failure is caused by a subtle effect of competing orbital states. Such orbital states are readily elucidated in a careful DFT-based study, and reveal an unusually weak coupling to the magnetism.

Our DFT calculations are performed in a full-potential code with a local-orbital basis set (FPLO)~\cite{fplo}. We used the experimental crystal structure, the local density approximation (LDA) with the exchange-correlation potential by Perdew and Wang~\cite{pw92}, and well-converged $k$ meshes of $1500-2000$ points in the symmetry-irreducible parts of the first Brillouin zone. The application of a generalized-gradient-approximation (GGA) exchange-correlation potential led to quantitatively similar results.

\begin{figure}
\includegraphics{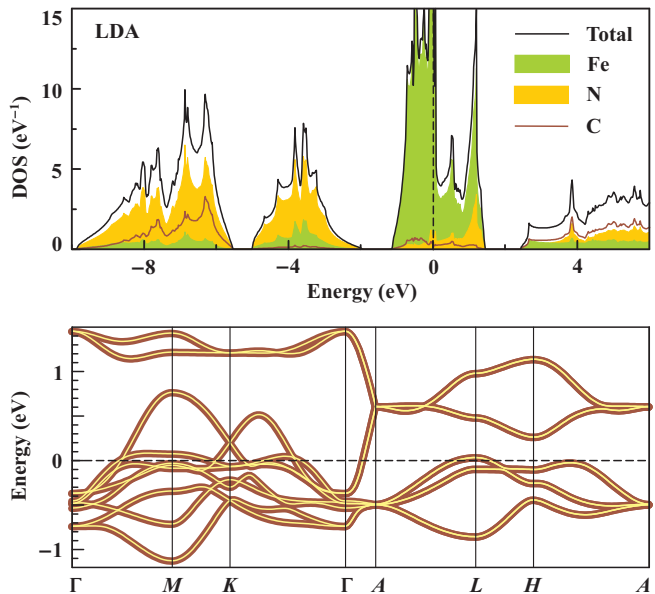}
\caption{\label{fig:band}
(Color online) Top: LDA density of states for FeNCN. Bottom: Band structure (thin light lines) and the fit with the tight-binding model (thick dark lines). The Fermi level is at zero energy.
}
\end{figure}

The LDA energy spectrum for FeNCN (Fig.~\ref{fig:band}) strongly resembles that of iron oxides. Nitrogen $2p$ states form valence bands below $-2$~eV, whereas Fe $3d$ states are found in the vicinity of the Fermi level. The energy spectrum is metallic due to the severe underestimation of electronic correlations in LDA.

The local picture of the electronic structure stems from the crystal-field levels of Fe$^{+2}$ with the electronic configuration $d^6$. The octahedral local environment induces the conventional splitting of five $d$ states into $t_{2g}$ and $e_g$ levels. This primary effect is accompanied by a weak trigonal distortion that further splits the $t_{2g}$ states into the $a_{1g}$ singlet and $e_g'$ doublet (Fig.~\ref{fig:str}, bottom). The balance between the $a_{1g}$ and $e_g'$ states is determined by fine features of the local environment. According to simple electrostatic arguments, the squeezing of the octahedron along the three-fold axis should slightly favor the $a_{1g}$ state (note the N--Fe--N angles $\alpha=95.9^{\circ}$ and $\beta=84.1^{\circ}$ in Fig.~\ref{fig:str}), which is represented by $d_{3z^2-r^2}$ orbital in the conventional coordinate system ($z$ along the $c$ axis). 

\begin{figure}
\includegraphics{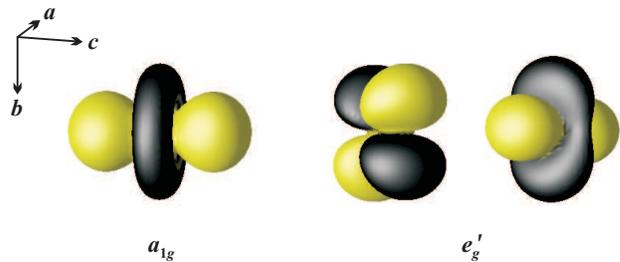}
\caption{\label{fig:wannier}
(Color online) Wannier functions based on the $a_{1g}$ and $e_g'$ orbitals.
}
\end{figure}

To quantify the orbital energies, we fit the Fe $3d$ bands with a tight-binding model based on Wannier functions adapted to specific orbital symmetries (Fig.~\ref{fig:wannier}). The fit yields $\eps_{a_{1g}}=-0.29$~eV, $\eps_{e_g'}=-0.30$~eV, and $\eps_{e_g}=0.58$~eV. The $t_{2g}-e_g$ splitting of about 0.9~eV is typical for $3d$ systems, whereas the energy separation of 10~meV between the $a_{1g}$ and $e_g'$ states is very small and opposite to the naive crystal-field picture. The difference may arise from covalency effects and/or long-range interactions inherent to solids (note a similar example in Ref.~\cite{deepa2008}).

Similar to iron oxides, FeNCN is expected to feature the high-spin state of Fe$^{+2}$ (the non-magnetic low-spin state apparently contradicts the experimental magnetic response reported in~\cite{liu2009}). Therefore, in the local picture five out of six $d$ electrons occupy each of the $d$ orbitals, whereas the sixth electron takes any of the $a_{1g}$ or $e_g'$ orbitals, thereby creating orbital degrees of freedom. Since the Mott-insulating state implies integer orbital occupations, the ground state of FeNCN should feature one doubly-occupied and four singly-occupied orbitals. The nature of the doubly occupied orbital is determined by the energy difference between $a_{1g}$ and $e_g'$ and, more importantly, by correlation effects.

To account for correlation effects in FeNCN, we use the DFT+$U$ method that treats strong electronic correlations in a mean-field approximation valid for insulators. In contrast to Ref.~\cite{xiang2010} reporting the half-metallic solution, we readily obtained insulating solutions by explicitly considering orbitally-ordered configurations. The failure of the previous computational work is likely related to spurious solutions arising from random starting configurations. Such solutions are not converged in charge due to the charge shuffling effect in a (half)-metal. To overcome this problem, we first performed calculations with a fixed occupation matrix (i.e., fixed the orbital configuration) and later released this matrix to allow for a fully self-consistent procedure. A similar approach has been used in previous computational studies~\cite{deepa-mno,*deepa2007,*janson2010,deepa2008}, and was shown to be vital for the proper treatment of systems with orbital degrees of freedom.

The input parameters of the DFT+$U$ method, the on-site Coulomb repulsion ($U_d$) and exchange ($J_d$), are evaluated in a constrained LDA procedure~\cite{constrained} implemented in the TB-LMTO-ASA code~\cite{lmto}. We find $U_d^{\LMTO}=6.9$~eV and $J_d=0.9$~eV, whereas a comparative calculation for FeO yields similar values of $U_d^{\LMTO}=7.1$~eV and $J_d=0.9$~eV. The magnitude of electronic correlations in FeNCN is, therefore, the same as in Fe$^{+2}$ oxides. By contrast, a somewhat reduced $U_d$ parameter was found in CuNCN (6.6~eV vs. $9-10$~eV in Cu$^{+2}$ oxides) and ascribed to sizable hybridization between the Cu $3d$ and N $2p$ states~\cite{tsirlin2010}. In FeNCN, such a hybridization is rather weak (Fig.~\ref{fig:band}).

Since the two Fe sites in the unit cell of FeNCN are crystalographically equivalent, we restrict ourselves to ferro-type orbital configurations featuring the same doubly occupied orbital on both Fe sites. Starting from different occupation matrices, we were able to stabilize several solutions~%
\footnote{We use the ground-state antiferromagnetic spin configuration with parallel spins in the $ab$ plane and antiparallel spins in the neighboring layers.}. %
The ground-state configuration features the doubly occupied $a_{1g}$ orbital and is further referred as $A_{1g}$. The $E_g'$ configurations with two electrons on either of the $e_g'$ orbitals lie higher in energy for about 75~meV/f.u. This energy difference is nearly independent of the specific $U_d$ value. It is also possible to put two electrons on one of the $e_g$ orbitals, but such configurations are highly unfavorable (0.81~eV/f.u. above the ground state) in agreement with the large $t_{2g}-e_g$ splitting of 0.9~eV in LDA. The artificial low-spin configuration with all six electrons on the $t_{2g}$ orbitals has an even higher energy of 3.6~eV/f.u. above the $A_{1g}$ ground state.

The lowest-energy spin configurations are antiferromagnetic, irrespective of the $A_{1g}$ or $E_g'$ orbital state. The energy spectra are quite similar~%
\footnote{See Supplementary information for representative energy spectra calculated with DFT+$U$.},
although the band gap ($E_{\gap}$) for the $A_{1g}$ orbital configuration is systematically higher than for any of the $E_g'$ configurations: for example, at $U_d=7$~eV $E_{\gap}=2.95$~eV and $2.30-2.50$~eV for $A_{1g}$ and $E_g'$, respectively. The change in $U_d$ causes a systematic shift of the band gaps~
\footnote{Compare to $E_{\gap}=2.45$~eV and $E_{\gap}=1.55-1.85$~eV for the $A_{1g}$ and $E_g'$ configurations, respectively, at $U_d=5$~eV.}. While the lack of the experimental optical data prevents us from tuning the $U_d$ parameter against the experimental band gap, we note that the calculated $E_{\gap}$ values agree well with the dark-red color of FeNCN.

We now investigate the interplay between spin and orbital degrees of freedom in FeNCN. To evaluate magnetic couplings, we doubled the unit cell in the $ab$ plane, and calculated total energies for several spin configurations. These total energies were further mapped onto the classical Heisenberg model yielding individual exchange integrals $J_i$. We evaluated the nearest-neighbor coupling $J_{ab}$ in the $ab$ plane as well as the nearest-neighbor and next-nearest-neighbor interplane couplings $J_c$ and $J_c'$, respectively (Fig.~\ref{fig:str}). Further couplings are expected to be weak due to negligible long-range hoppings in our tight-binding model.

\begin{table}
\caption{\label{tab:exchange}
Exchange couplings (in~K) calculated for the ground-state ($A_{1g}$) and one of the higher-lying ($E_g'$) configurations. Interatomic distances are given in~\r A (see also Fig.~\ref{fig:str}). The on-site Coulomb repulsion parameter is $U_d=7$~eV.
}
\begin{ruledtabular}
\begin{tabular}{rcrr}
           & Distance & $A_{1g}$ & $E_g'$ \\\hline
  $J_{ab}$ &   3.27   &    $-11$ & $-6$   \\
  $J_c$    &   4.70   &       51 &   58   \\
  $J_c'$   &   5.73   &        2 &   0    \\
\end{tabular}
\end{ruledtabular}
\end{table}

Surprisingly, the calculated exchange couplings listed in Table~\ref{tab:exchange} depend only weakly on the orbital order. Both $A_{1g}$ and $E_g'$ orbital configurations induce leading antiferromagnetic (AFM) exchange $J_c$ via the NCN groups. The intraplane coupling $J_{ab}$ is ferromagnetic (FM), whereas $J_c'$ is AFM. The resulting spin lattice is non-frustrated, and features AFM long-range order with parallel spins in the $ab$ plane and antiparallel spins in the neighboring planes. This prediction awaits its verification by a neutron scattering experiment. To test our microscopic magnetic model against the available experimental data, we simulated the magnetic susceptibility using the quantum Monte-Carlo \texttt{loop} algorithm~\cite{loop} implemented in the ALPS package~\cite{alps}. The susceptibility was calculated for a three-dimensional $L\times L\times L$ finite lattice with periodic boundary conditions and $L=12$~\footnote{$L=12$ is sufficient to eliminate finite-size effects for the magnetic susceptibility within the temperature range under consideration.}.

\begin{figure}
\includegraphics{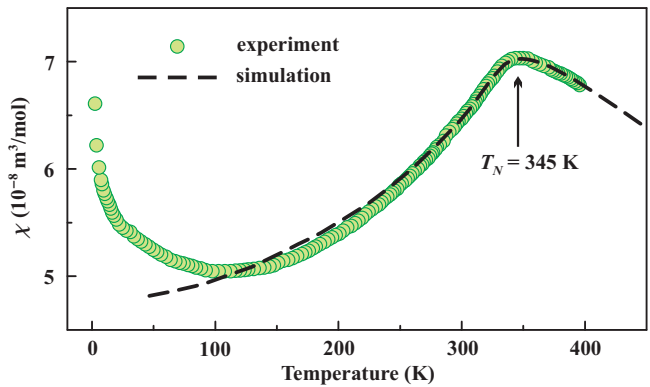}
\caption{\label{fig:chi}
(Color online) Experimental magnetic susceptibility of FeNCN~\cite{liu2009} and the simulated curve for $J_c=46.5$~K, $J_{ab}=-16.3$~K, and $J_c'=2.3$~K. The deviations at low temperatures are due to an impurity contribution and/or anisotropy effects.
}
\end{figure}

The simulated magnetic susceptibility was compared to the experimental data from Ref.~\cite{liu2009} (Fig.~\ref{fig:chi}). The data above 100~K and, particularly, the transition anomaly at $T_N=345$~K are perfectly reproduced with $J_c=46.5$~K, $J_{ab}=-16.3$~K, and $J_c'=2.3$~K in remarkable agreement with the calculated exchange couplings for the $A_{1g}$ orbital configuration (Table~\ref{tab:exchange})~%
\footnote{The fitted $g$-value is $g=1.98$.}%
. The low-temperature upturn of the experimental susceptibility violates the behavior expected for a Heisenberg antiferromagnet, and signifies an impurity contribution or effects of exchange anisotropy that are not considered in our minimum microscopic model. 

The three-dimensional magnetism of FeNCN is somewhat unexpected considering the seemingly layered nature of the crystal structure (Fig.~\ref{fig:str}). The leading exchange is antiferromagnetic and runs between the layers, although the nearest-neighbor interlayer distance of 4.70~\r A is much longer than the intralayer distance of 3.27~\r A. The unusually strong interlayer exchange originates from the peculiar nature of the NCN units that feature a strong $\pi$-bonding and mediate hoppings between the neighboring layers. This effect has been illustrated by sizable contributions of both nearest-neighbor and second-neighbor nitrogen atoms to Wannier functions in CuNCN~\cite{tsirlin2010}. Similar contributions are found for the $e_g$ orbitals in FeNCN. 

The intralayer interaction is a conventional combination of the direct exchange and Fe--N--Fe superexchange that result in a weakly ferromagnetic coupling (the Fe--N--Fe angles are $95.9^{\circ}$, i.e., close to $90^{\circ}$). The spin lattice of FeNCN reminds of another transition-metal carbodiimide, CuNCN, where the long-range antiferromagnetic superexchange mediated by the NCN groups was also reported~\cite{tsirlin2010}. The difference between the two compounds is the strong Jahn-Teller distortion in CuNCN that splits the close-packed layers of transition-metal octahedra into structural chains running along $a$, with the ferromagnetic coupling resembling $J_{ab}$ in FeNCN. By contrast, FeNCN is not subjected to a Jahn-Teller distortion, and the ground-state orbital configuration is largely stabilized by electronic correlations. Indeed, there is no clear preference for a certain orbital state on the LDA level.

According to our results, the energy scales for the spin and orbital degrees of freedom in FeNCN are comparable, about 52~meV/f.u. and 75~meV/f.u., respectively. However, magnetic couplings weakly depend on the orbital configuration keeping spins and orbitals nearly decoupled. This effect is explained by different $d$ states responsible for the orbital and magnetic effects. The orbital degrees of freedom are operative in the $t_{2g}$ subspace, whereas magnetic couplings are largely determined by the $e_g$ orbitals featuring stronger intersite hoppings. The decoupling of spin and orbital variables along with the low energy scale of the competing orbital states suggest that the orbital-only physics can be probed in FeNCN.

The characteristic energy scale of 75~meV/f.u. corresponds to temperatures around 850~K, and implies that the $E_g'$ orbital states may emerge at elevated temperatures. Although FeNCN, alike all transition-metal carbodiimides, is thermodynamically unstable~\cite{launay2005}, it can be maintained up to at least 680~K, which is the preparation temperature reported in Ref.~\cite{liu2009}. Other options of activating the $E_g'$ orbital states of FeNCN are the application of pressure and laser irradiation. The latter has been successfully used for melting the orbital order in several prototype orbital systems~\cite{tomimoto2003,*mazurenko2008} and could be applied to FeNCN as well. If the switching of the orbital state is possible, the anticipated effect is a sizable reduction in the band gap (for at least 0.4~eV), while the structure will probably adjust to the new orbital state. However, it is more likely that several competing $E_g'$ states will form an orbital liquid, thereby maintaining the high symmetry of the crystal structure. To probe such effects, further experimental work on FeNCN is highly desirable. We also mention an isostructural compound CoNCN~\cite{krott2007}, where orbital degrees of freedom arising from Co$^{+2}$ ($d^7$) are expected. 

In summary, we have shown that FeNCN presents an unusual example of weakly coupled spin and orbital degrees of freedom acting on a similar energy scale. The ground-state orbital configuration features two electrons on the $a_{1g}$ orbital, in agreement with the simple electrostatic arguments, but in contrast to the LDA-based expectations. The calculated properties, such as the band gap, exchange couplings, and N\'eel temperature, are in very good agreement with the experiment. We have also remedied the failure of the recent computational work~\cite{xiang2010} and confirmed the remarkable performance of DFT+$U$ techniques applied to Mott insulators with orbital degrees of freedom. 

We are grateful to Deepa Kasinathan and Oleg Janson for fruitful discussions. We also acknowledge Richard Dronskowski and Andrey Tchougr\'eeff for drawing our attention to FeNCN. A.T. was funded by Alexander von Humboldt Foundation.

%

\begin{widetext}
\begin{center}
{\large
Supplementary information for 
\smallskip

\textbf{``Orbital order and magnetism of FeNCN''}}
\medskip

Alexander A. Tsirlin, Klaus Koepernik, and Helge Rosner
\end{center}
\medskip

\begin{figure}[!h]
\includegraphics{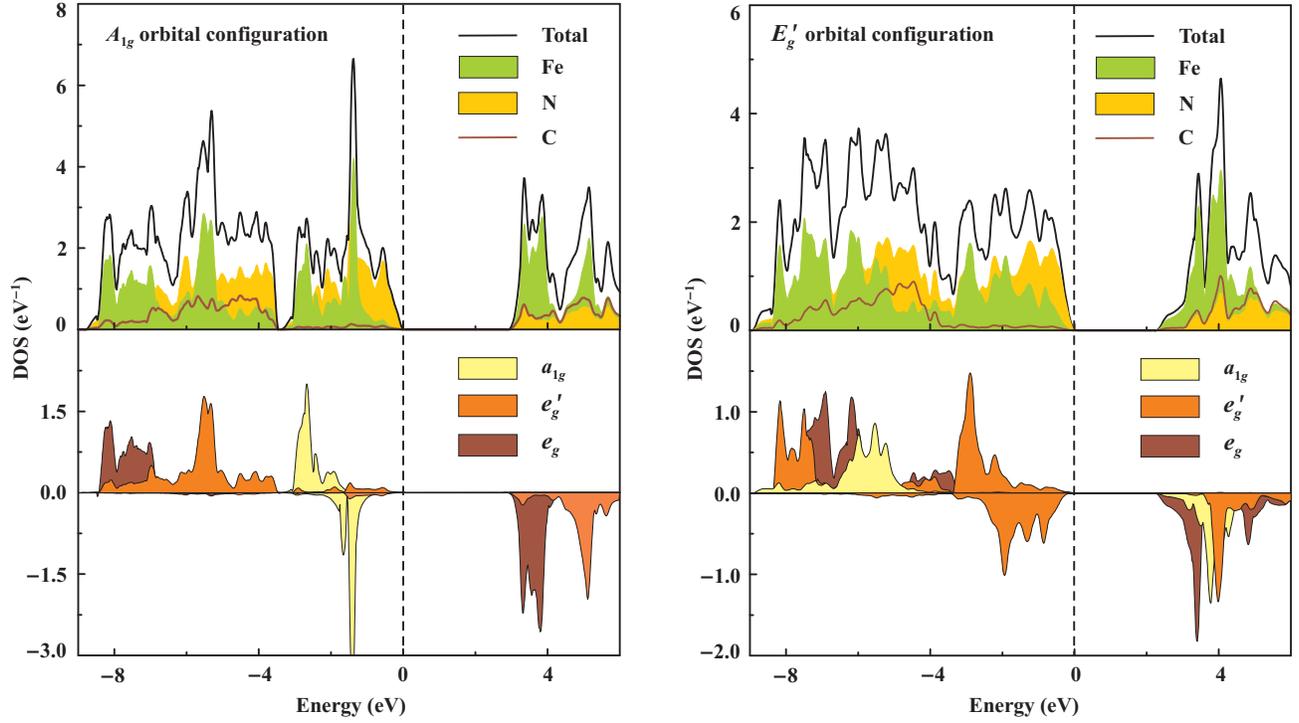}
\caption{\label{fig:s1}\normalsize
Atomic- and orbital-resolved density of states for the $A_{1g}$ (left panel) and one of the $E_g'$ (right panel) orbital configurations of FeNCN. The Fermi level is at zero energy. The on-site Coulomb repulsion parameter is $U_d=7$~eV.
}
\end{figure}
\end{widetext}

\end{document}